\documentclass[11pt,twoside]{article}
\usepackage{amsmath}
\usepackage{asp2010}

\resetcounters

\begin{document}

\title{Recent results in Ring Diagram analysis}
\author{M.~Cristina~Rabello-Soares 
\affil{Department of Physics, Universidade Federal de Minas Gerais, Belo Horizonte, MG 31270, Brazil}
}

\begin{abstract}

The ring-diagram technique was developed by Frank Hill 25 years ago and developed quickly during the late 1990s. 
It is nowadays one of the most commonly used techniques in local helioseismology. 
The method consists in the power spectral analysis of solar acoustic oscillations on small regions (2 to 30 degrees) of the solar surface. 
The power spectrum resembles a set of trumpets nested inside each other and, for a given frequency, it looks like a ring, hence the technique's name. 
It provides information on the horizontal flow field and thermodynamic structure in the layers immediately below the photosphere. 
With data regularly provided by MDI (on board SOHO), GONG+ network and more recently HMI (on SDO), many important results have been achieved. 
In recently years, these results include estimations of the meridional circulation and its evolution with solar cycle; flows associated with active regions, as well as, flow divergence and vorticity; and thermal structure beneath and around active regions. 
Much progress is expected with data now provided by HMI's high spatial resolution observations and high duty cycle. 
There are two data processing pipelines (GONG and HMI) providing free access to the data and the results of the ring-diagram analysis. 
Here we will discuss the most recent results and improvements in the technique, as well as, the many challenges that still remain.

\end{abstract}

\section{Introduction}

The two-dimensional power spectra ($k - \omega$) of solar oscillations have
discrete ridges of power, which are given by the dispersion relation between the temporal frequency $\omega$ and total horizontal wavenumber $k$ for acoustic modes with a given radial order $n$. 
For modes with spherical harmonic degree $\ell$ larger than 190, 
the ridges appear as continuous structures rather than a series of individual peaks (each corresponding to an integer $\ell$), indicating that a local plane-wave representation rather than a global spherical harmonic decomposition can be used to analyze the oscillations \citep{MCRS_patron1995, MCRS_gonzalez1999}, specially in a small region of the solar surface where its curvature can be neglected.
For these high-degree modes, we can compute three-dimensional spectra as function of two horizontal wavenumber components, $k_x$ and $k_y$, and $\omega$.
The power spectrum 
resembles a set of trumpets nested inside each other and, for a given frequency, 
it looks like a set of concentric rings, hence the technique's name \citep{MCRS_hill1988}.
Each ring corresponds to a particular value of the radial order $n$.
In the absence of flows, 
these trumpets would have their axis of symmetry coinciding with the $\omega$ axis, and rings would be circles. 
Since the wave fronts are being advected (specially, by solar rotation) the ridges are thus tilted and the rings are distorted into shapes close to ellipses.
It thus provides information not only on the thermodynamic structure (through the dispersion relation determination), but also on the horizontal flow field in the layers immediately below the photosphere.

It is common to track a small region of Sun (usually 15\deg\ aperture, but also apertures from 2\deg\ to 30\deg\ are used) to eliminate the large advection due to solar differential rotation. 
The tracking time is usually the time it takes a region to rotate through its diameter, about 27 hours for the 15\deg\ patch.
The image patches are then remapped into solar coordinates and apodized before being Fourier transformed in three dimensions
\citep[see, for example,][]{MCRS_bogart1995, MCRS_haber2002}.
The remapping often utilizes Postel's projection, which preserves the distances along great circles \citep{MCRS_haber2002}.
The spectra are then fitted to a ring model which estimates (as fit parameters)
amplitudes, widths and central frequency as a function of $k$ for each $n$ value.
Assuming a simple power law as the dispersion relation for the waves in the absence of a flow, $\omega_0 = c k^p$, 
which provides information on the thermodynamics of the material below the photosphere.
The presence of a velocity field {\bf $U$} produces an apparent Doppler shift $\Delta\omega$:
$\omega = \omega_0 + \Delta\omega = c k^p + k_x U_x + k_y U_y$, 
where $U_x$ and $U_y$ are the components of the horizontal velocity in the zonal and meridional directions, respectively
\citep{MCRS_hill1988}.
The final step is to invert the estimated mode velocities 
to obtain their dependence with depth 
and the mode frequencies to obtain the dependence with depth of the thermal structure,
in particular sound speed \citep[see][for more details]{MCRS_ab2007}.
For the typical 15\deg\ diameter region, the inferences extend down to about 16 Mm below the solar surface.

With data regularly provided by the Michelson Doppler Imager \citep[MDI:][]{MCRS_scherrer1995} on board the Solar and Heliospheric Observatory (SOHO), 
the Global Oscillation Network Group \citep[GONG+:][]{MCRS_harvey1996, MCRS_harvey1998} 
and most recently Helioseismic and Magnetic Imager {\citep[HMI:][]{MCRS_schou2012} on the Solar Dynamics Observatory (SDO), many important results have been achieved.
GONG data are available with sufficient spatial resolution for ring analysis since 2001.
HMI, launch in the beginning of 2010, provides almost continuous high spatial resolution observations.
Although MDI (launch in December 1995) was turned off in April 2011, its 15-year data set is still has valuable information
and, after cross-calibration, can be used to extend HMI observations into cycle 23.

There are two data processing pipelines 
\citep[HMI and GONG:][respectively]{MCRS_bogart2011a, MCRS_bogart2011b, MCRS_corbard2003}
providing free access to the data and the results of the ring-diagram analysis.
Each pipeline analyses regularly its own data (in the case of HMI pipeline both HMI and MDI data are analysed), but they are also capable of analyse each other data and, in fact, it is often done.
The HMI 
pipeline uses two fitting models to the power spectrum: `rdfitc' \citep{MCRS_ba1999} 
and `rdfitf' \citep{MCRS_haber2002}. 
The later is also implemented in the GONG pipeline.
Both methods use a Lorentzian profile. 
However, among other differences, in the case of `rdfitc', there is the addition of an asymmetry term.

Magnetic fields play an importan role in many astrophysical processes from the Sun to the Universe as a whole.
However, vector magnetic fields are not easily measured and, 
since the relevant equations are highly non-lineal, they are also difficult to model, hence it is properly dubbed as `the elephant in the room' (from the 25th annual Canary Islands Winter School of Astrophysics 2013 on cosmic magnetic fields).
Thus, despite the effort, 
the origin and structure of sunspots is still a matter of debate, 
with different proposed models.
The interactions between sunspots and solar acoustic modes could be used to probe the sub-surface structure of sunspots \citep[see, for example,][and references within]{MCRS_gizon2010r}.
However, the
inadequate understanding and modelling of the interactions between the acoustic waves and the sunspots, where magnetohydrodynamic effects dominate,
has led to apparently contradictory results reported in the literature \citep[e.g.][]{MCRS_rajaguru2011}.
This is an important and difficult problem.
Therefore, a significant effort has been and continues to be expended in it. 
Thus, it will be discussed in all sections. 

This paper describes recent results using Ring Diagram analysis
and it is divided as follows.
We present results on:
global flows (Section 2),
local flows associated with active regions (Section 3),
solar flares (Section 4),
thermodynamic structure variation (Section 5),
acoustic wave parameters (Section 6) and
observations at different heights in the solar atmosphere (Section 7). Finally, in Section 8, we present our final remarks.
There is also another good review on recent results by \citet{MCRS_baldner2012}.


\section{Global Flows}

\subsection{Meridional flow}

\citet{MCRS_komm2012} using the HMI ring-diagram pipeline for 18 Carrington rotations (CR 2097 to 2114) obtained a maximum amplitude for the average meridional flow of about 20 m/s near 37.5\deg\ latitude.
This is comparable to previous ring-analysis results obtained using GONG and MDI data \citep{MCRS_gonzalez2008, MCRS_ba2010} 
and, also, to results obtained at the solar surface from Mt. Wilson Observatory (MWO) Doppler measurements \citep{MCRS_ulrich2010}.

However, tracking of magnetic features at the solar surface leads to meridional flows with $\approx$~40\% smaller amplitudes \citep{MCRS_hataway2010}.  
Since the measured meridional flow varies with disk position (and, thus, different geometric foreshortening), 
\citet{MCRS_zhao2012} used the observed east-west variation of the zonal flow 
as a measurement of a north-south systematic variation present in the meridional flow estimation.
The resulting meridional flow has a better agreement with those obtained from tracking of magnetic features. 
\citet{MCRS_baldnerschou2012} 
analysed the effect of the vertical flows from convection in the outer solar convection zone and atmosphere and concluded that the ad hoc correction applied by \citet{MCRS_zhao2012} is likely justified.
Another controversial result is 
that the amplitude of the meridional flow does not decrease with depth, which is expected since the density increases with depth
\citep[see][and references within]{MCRS_komm2012}. 
For more recent results on meridional circulation variation with depth, see
the paper by \citet{MCRS_greer_this} in this proceddings.

Another question is whether the meridional flow remains poleward or reverses direction at some high latitude (the so-called `counter cells').
HMI high-resolution data provide better coverage at high latitudes than
MDI or GONG data, which are limited to $\pm$52.5\deg\ latitude.
Using HMI data and
taking into account systematic effects due to $B_0$-angle variation,
\citet{MCRS_komm2012} found that
the meridional flow is poleward at all depths within $\pm$~67.5\deg latitude
and
the flow is equatorward only at $75^\circ$ latitude, 
indicating a countercell very close to the poles during cycle 23.

It has been observed that the meridional circulation varies with time,
where
the amplitude of the meridional circulation decreases with solar activity 
\citep[see][and references within]{MCRS_gonzalez2010}.
More recently,
\citet{MCRS_ba2010} 
expanded the meridional flow, $u_y(r,\theta,t)$, in terms of associated Legendre polynomials of order 1,
$P^1_i (cos\theta)$ where $\theta$ is the angle from the north pole.
They show that the expansion coefficient $c_i(r,t)$ of degree $i=2$ is anticorrelated with solar cycle using MDI data from 1996 to 2009 (their Figure 6).

\subsection
{Correlation of meridional and zonal flows with sunspot butterfly diagram}

The correlation between the flow variation and the magnetic field strength 
as a function of latitude and time has been observed by several authors,
where
migrating patterns of the meridional and zonal flows match those of the sunspot butterfly diagram 
\citep{MCRS_gonzalez2010, MCRS_ba2011, MCRS_komm2011_a}.  
\citet{MCRS_gonzalez2010} also presented evidence of the appearance 
at mid-latitudes
of a meridional flow pattern migrating towards the poles
prior to the emergence of activity with solar cycle 24. 
Like 
the migrating zonal flow pattern known as the `torsional oscillation', which consists of belts of slightly faster than average rotation that migrate from mid-latitudes to the equator and poles.
\citet{MCRS_komm2011_a} confirmed these results observing
faster-than-average bands of the meridional and zonal flow associated with the new cycle.
For the zonal flow,
the migration with latitude of the flow pattern 
towards the equator during the declining phase of the solar cycle
is apparent in the deeper layers (up to 16 Mm), while for the meridional flow, only in the layers close to the surface \citep[see Figure 1 in][]{MCRS_komm2011_a}. 
Results at deeper layers (35 Mm) for the zonal flow are obtained by global helioseismology (using Spherical Harmonic Decomposition)
and agree with ring-analysis results \citep[see][]{MCRS_howe2013a}.
Furthermore,
\citet{MCRS_komm2011_a} shown that
the north-south difference pattern of the zonal flow resembles the cycle variation of the meridional flow (their Fig. 2),
bringing up the question of whether the two patterns are connected.
For more recent results on large-scale flows see the papers
by \citet{MCRS_bogart_this, MCRS_komm_this}
in this proceddings.


\section{Local flows associated with active regions}

\subsection{Emergence and decay of active regions}

\citet{MCRS_komm2011_hh} 
studied the temporal variation of subsurface flows of more than 800 active regions using 
GONG and some MDI data for comparison.
The smoothed variation of the unsigned magnetic flux at solar surface
were used to characterize emergence and decay of the active regions.
During the active region emergence, they observed faster-than-average zonal subsurface flow while, during decaying, it the zonal flow
slows down to values comparable to that of quiet regions (their Figures 10 to 12).
The meridional velocity is
mainly poleward and shows no obvious variation. 
They also estimated the vertical velocity
from the divergence of the measured horizontal flows taking into account mass conservation
and some of their main results are the following.
During emergence, there are upflows between 10 and 16 Mm in depth.
At established active regions, there are
downflows in shallow layers ($4-10$ Mm) and upflows at depths greater 
than $\approx$~10 Mm.
At decaying active regions, upflows become weaker at deeper layers. 
See also the paper by \citet{MCRS_jain_this_flow} in this proceddings.

\subsection{In- and outflows}

\citet{MCRS_hht2009} 
using 2\deg\ patches 
(which are much smaller than the standard 15\deg\ size)
observed a mean surface inflow of approximately 20 m/s (at depths of 2 Mm or less).
And, from results obtained through a variety of techniques by different authors,
they proposed the following schematic diagram
(their Figure 11).
The surface cooling within the plage surrounding the sunspot results in a downdraft which draws fluid in at the surface (i.e., a inflow from the quiet region towards the plage) which is
followed by 
an outflow at deeper layers ($>$~10 Mm).
In constrast, 
closer to the active region center, 
the downflow within the plage,
results in a moat outflow 
(from the sunspot towards the plage)
at shallow layers ($<$~2 Mm deep)
and a inflow at slightly deeper layers.
In a recent review, \citet{MCRS_gough2012} pointed out that the qualitative picture, which has made its way onto the well-known SOI/MDI coffee mug
\citep[Figure 6 of][and references within]{MCRS_kosovichev2012}.
is now not only physically plausible, but also in accord with photospheric observations.
The deep divergent horizontal flow carries away the excess heat rising around the obstructing sunspot and an associated convergent subphotospheric above.

\subsection{Multi-scale inversion}

\citet{MCRS_fht2011} 
developed a multi-scale 3-D inversion procedure using three different region sizes (2\deg\, 4\deg\, and 16\deg)
and applied to MDI data taken during CR 1985 (January 2002).
They found subsurface outflows around sunspots 
to a depth of about 7 Mm.
Analysing a given sunspot in their data set, they observed
that outflows are visible at all depths, but decay sharply below 7 Mm; and
that there are two outflow components: one which has its largest value around 1 Mm and another around 5 Mm
(their Figure 3a).
The superficial flow ($\approx$~1 Mm) decays more sharply with depth
than the deeper one ($\approx$~5 Mm).
The authors pointed out that the two components
may be driven by different physics which might affect time-distance and ring-analysis differently.
For more recent results on this, see \citet{MCRS_hindman_this}.

\subsection{Rotating flows}

As mentioned in Section (3.2), there is a inflow from the surrounding quiet regions towards the active region and a moat outflow close to its center.
Coriolis forces act upon these surface flows and spin up the flow.
\citet{MCRS_hht2009} 
using high-resolution ring diagrams (2\deg\ patches) 
found cyclonic convergent flows at the periphery of the active region
and
anti-cyclonic divergent flows in its core
(their Figure 12 shows a schematic diagram).

To measure the flow rotation, one can calculate the vorticity:
$\boldsymbol{\omega} = \boldsymbol{\nabla} \times \boldsymbol{v}$ ,
where $\boldsymbol{v}$ is the vector flow
\citep[e.g.,][]{MCRS_komm2004}.
Besides the vertical vorticity from the near surface cyclonic flows, 
there is also a horizontal vorticity 
(with two components, meridional and zonal vorticity)
from 
the flow circulation in the active region, given by 
the up- and downflows coupled with in- and out flows.
\citet{MCRS_khh2012} 
estimated the vorticity temporal variation
of 
more than 800 active regions using GONG data (15\deg\ regions),
subtracting the mean at every depth.
The vertical vorticity seems to be constant with time.
While the horizontal vorticity 
increases during 
the active region emergence and decreases during its decay
(i.e., during an increase or decrease in the unsigned magnetic flux at solar surface, respectively).
There are some indications that the increase in horizontal vorticity during 
emergence happens about a day later at depths below about 8 Mm compared to shallower layers and a day earlier during 
decay (their Figure 11).

Besides the observed surface cyclonic flows, some sunspots are known to rotate around their umbral center \citep[etc.]{MCRS_evershed1909, MCRS_brown2003}.
\citet{MCRS_jain2012} 
studied a fast-rotating sunspot (NOAA 10930 active region) 
using $11^\circ$ regions.
The flow associated with the rotating sunspot varies significantly with depth during the course of the rotation while the flow in non-rotating sunspots are constant. 
They also found evidence of two opposite flows at different depth
indicating a twist in the magnetic field lines.


\section{Solar flares}

It is vital to space weather that we improve our understanding of solar flares 
and coronal mass ejections (CMEs) 
and increase our ability to forecast when they will occur and how large they will be.
Highly twisted magnetic fields are very likely responsible for strong eruptive phenomena such as flares and CMEs 
and measured vorticity of subsurface flows might be useful to 
space weather forecast 
\citep[and references within]{MCRS_reinard2010}.
\citet{MCRS_komm2011_b}
estimated the vorticity and helicity below more than 1000 active regions 
(using GONG data)
and concluded that 
they
improve the ability to distinguish between flaring and non-flaring active regions.


\citet{MCRS_wolff1972} 
first suggested that energetic flares may be able to excite acoustic waves by exerting mechanical impulse of the thermal expansion of the flare on the photosphere. 
Although excitation of flare-related traveling waves on the solar surface have been reported \citep{MCRS_kz1998}, 
this is not clear for 
standing waves or normal modes of solar oscillations.
\citet{MCRS_mat2009} 
argued that 
any mode amplification induced by transient flare events has to compete with the absorption effects associated with intense magnetic fields of sunspots
and might not be resolved by averaging techniques.
They 
found strong evidence that the
mode amplitude increased significantly after 
the long duration energetic X17.2/4B flare of October 28, 2003
(associated to NOAA 10486 active region) 
and found systematic variations in sub-surface flows.
For more recent results, see the paper by \citet{MCRS_maurya_this} in this proceddings.

\citet{MCRS_howe2013b} 
did not find any systematic differences 
in mode amplitude and frequency between coronal holes and other quiet-Sun regions.


\section{Thermodynamic structure variation}

As initially reported by \citet{MCRS_wn1985}, 
it is now well established that the frequency of the solar acoustic modes is 
strong correlated with solar activity
and also with the 
magnitude of the local surface magnetic field 
\citep[e.g.][]{MCRS_hindman2001}.  
The observed mode frequencies can be inverted to obtain the 
thermal structure (in particular, sound speed) 
in the solar interior where the waves travel through. 

\citet{MCRS_bbb2012} 
analysed 264 active regions (from 1996 to 2008) using MDI data and 
applying principal component analysis.
They confirmed previous results 
\citep{MCRS_bab2004, MCRS_bogart2008}, 
a distinctive two-layer thermal structure beneath surface magnetic activity,
with a shallow negative sound speed perturbation and a somewhat deeper positive perturbation (their Figure 9);
and a
similar structure for the adiabatic index, $\Gamma_1$ (their Figure 11).
The size of these perturbations increases with
the magnitude of the surface magnetic field
before reaching saturation.

However,
\citet{MCRS_llb2006} 
showed that `sound speed' obtained by inverting frequency differences between active and quiet regions of the Sun 
is in fact a combination of both sound speed and Alfv\'en speed, 
which they called `wave speed':
$c_T^2 \equiv \Gamma_1 P_T / \rho$ where 
$P_T$ and $\rho$ are the total pressure (gas pressure, $P_{gas}$, plus magnetic pressure, $P_{mag}$) and density, respectively.
\citet{MCRS_lbl2009} 
using solar models with magnetic fields have determined a relationship between 
$\delta\Gamma_1/\Gamma_1$ and $\delta\beta$ where $\beta \equiv P_{mag}/P_{gas}$.
Applying it to all regions studied by 
\citet{MCRS_bab2004}, 
they found that 
$\delta\beta$ has the same two-layer structure beneath 
the active region, but reversed, i.e., it 
is positive immediately below the surface 
but negative at deeper regions.

\citet{MCRS_gizon2009, MCRS_gizon2010e},
analysing NOAA 9787 active region,
shown that
there is a strong disagreement between 
different local-helioseismic methods, especially between the time-distance and ring-diagram results \citep[see Figure 16 of][]{MCRS_gizon2010r}. 
There are a number of possible explanations for this disagreement,
namely the possible naivety of modeling potentially complicated effects of the magnetic field in terms of an equivalent sound-speed perturbation
\citep[as shown, for instance, by][]{MCRS_lbl2009}.
Current inversions for the time-distance and ring-diagram methods use sensitivity functions that do not explicitly include the direct effects of the magnetic field, and both assume that wave-speed perturbations are small. 
Different
models estimate
a strong perturbation 
in the first 2 Mm below the surface \citep[e.g.,][]{MCRS_rempel2009, MCRS_fbc1995}
and, accordingly to \citet{MCRS_moradi2010},
only after properly addressing these near-surface effects will 
deeper inversions be reliable.

\citet{MCRS_rs2012}, 
using global helioseismology applied to MDI data (from 1999 to 2008),
presented evidence that the sound-speed 
variation with solar activity 
(i.e., variation between solar maximum and solar minimum)
has a two-layer configuration, similar to the one observed by ring analysis below an active region and at a similar location (her Fig. 9).
In addition,
performing a back of the envelope calculation, 
the size of the solar-cycle variation
is in agreement 
with an overall effet of the active regions present throughout the solar cycle.


\section{Acoustic wave parameters}

It is well known that amplitudes decrease while mode widths and frequencies increase in the presence of magnetic fields \citep[for example,][]{MCRS_rba2001}.
The mode amplitude variation rate 
is approximately a linear function of the width variation rate
\citep[Figure 4 of][]{MCRS_rbb2008}.

\citet{MCRS_burtseva2013} 
analysed 
variations in the amplitudes and widths 
during the declining phase of cycle 23 and rising phase of cycle 24 (from 2001 to 2013)
using GONG, MDI and HMI data. 
The mode parameters from all three instruments respond similarly to the varying magnetic activity. 
The mode amplitudes and widths (for $\ell$ = 440, $n$ = 2,  $\nu$= 3.2 mHz)
show consistently smaller variation 
in the rising phase of cycle 24 as compared with the declining phase of cycle 23 
due to the lower magnetic activity.
They also concluded that 
a long-term variation 
in the mode amplitude of quiet regions 
found from 2004 to 2008 \citep{MCRS_burtseva2011} 
is probably due to a problem in the data (their Figure 1).

With respect to the frequencies change,
\citet{MCRS_bbb2011} 
have found that the frequency changes 
are slightly different depending on the sunspot type.
And
\citet{MCRS_tjh2013} 
found that, 
during the extended minimum between the cycle 23 and 24,
both the spatial and temporal frequency shifts are weakly correlated with the surface magnetic field,
implying that the shifts can not be accounted by the regions of observed strong field component of the magnetic field alone. They conjecture that the changes in acoustic high-degree mode frequencies may be associated with two components: strongly localized active regions and the quiet-sun magnetic field. The former explains the larger shifts while the later may be responsible for the smaller shifts in all phases of solar activity.
For more recent results on this, see the paper by \citet{MCRS_tripathy_this} in this proceddings.

\citet{MCRS_rbs2013} 
instead of studying the wave characteristics in 
the presence of strong magnetic fields present in active regions,
looked at quiet regions close to an active region (less than 8\deg\ away).
They searched 
for perturbations in the wave characteristics as the solar acoustic oscillations are affected as they propagate inside the nearby sunspot,
using HMI 5\deg\ patches during 2.5 years of cycle 24 ascending phase.
They observed significant variations in the mode parameters and flows.
However,
the mode dependence of the variations does not have the same functional form as that observed for the differences between quiet and active regions.
They also analysed 
the dependence of variations
on the direction of the wave propagation.
In 
one of the fitting methods implemented in the HMI pipeline, `rdfitc', 
there is the addition of 
two terms for the azimuthal amplitude anisotropy:  
$a_2 \cos(2\theta) + a_3 \sin(2\theta)$.
For frequencies smaller than $\approx$4.2 mHz, the anisotropic variation is of the same order of magnitude as the constant term in the amplitude
and the modes are attenuated by as much as 20\%.
For larger frequencies, 
the modes are amplified (`acoustic halo') 
by as much as 20\%, if there is an active region nearby, 
with very little dependence on their propagation direction. 
There is flow variation only in the direction of the nearby active region. 

\section{Observations at different heights in the solar atmosphere} 

HMI provides different observables, 
which are sensitive to the five-minute acoustic oscillations,
that are formed at different heights in the photosphere \citep[see][]{MCRS_fleck2011, MCRS_norton2006},
such as:
Doppler velocity ($V$), continuum intensity ($I_C$), line depth ($L_d$), and line core ($I_L$).
While
the Atmospheric Imaging Array (AIA), also onboard SDO, 
observes at wavelengths 
that also
have good sensitivity to acoustic modes, specifically the 1600 and 1700 \AA\ bands
\citep{MCRS_howe2012}, 
and
are 
formed higher in the solar atmosphere,
i.e., 
in the lower chromosphere and upper photosphere, respectively 
\citep{MCRS_judge2001}.
By
analysing 
and comparing
the power spectra of the HMI and AIA observables,
it is possible to 
compare observations obtained at different heights in the solar atmosphere
to better understand the behaviour of the waves near (or past) its upper reflecting boundary.

\citet{MCRS_howe2012}
compared 
the characteristics of 
AIA and HMI observables spectra 
for a 15-degree patch around NOAA 11072 active region
They found that the
five-minute oscillation power in all observables is suppressed in the sunspot and also in plage areas. Above the acoustic cut-off frequency, the behaviour is more complicated: power in HMI $I_C$ is still suppressed in the presence of surface magnetic fields, while power in HMI $I_L$ and the AIA bands is suppressed in areas of surface field but enhanced in an extended area around the active region,
and power in HMI $V$ is enhanced in a narrow zone around strong-field concentrations and suppressed in a wider surrounding area
(their Figures 3 to 7).
These regions of enhanced power surrounding the magnetic field concentrations are commonly known as `acoustic halo'.
\citet{MCRS_tripathy2012},
following a similar method,
analysed the power spectra 
around NOAA 11092 active regin 
and found similar results (their Figure 6).
They also obtained that,
for 
different HMI observables ($V$, $I_C$, $L_d$),
zonal and meridional flows do not change significantly;
while the mode frequencies differ (their Figure 3).
For more recent results, see the papers by 
\citet{MCRS_tripathy_this, MCRS_jain_this_aia} 
in this proceddings.


\section{Final Remarks}

Much has been done using this technique and new results are continually reported.
An important element in
fuelling this progress is the huge quantity and high quality of helioseismic data being provided by HMI
in combination with
the free and almost immediately access to the analysis results
provided by the two data processing pipelines (GONG and HMI)
and, the fact, that 
we are now starting to have another cycle to compare with the well studied cycle 23.
As a consequence, there has been a need to improve the method.
Some of the improvements are described in the paper,
others are as follows.
The improvement of the inference of local flows 
by \citet{MCRS_birch2007} 
where they concluded that
the exact form of the sensitivity of ring measurements 
to actual flows in the Sun
depends on the details of the fitting procedure.
The concurrent fitting of multiple p-mode ridges,
instead of fitting each ridge individually,
by
\citet{MCRS_hindman_this}
which
improves not only the fitted parameter, but also increases the number of fitted modes.
Last but not least, the important work of
the validation of the ring-diagram technique 
through the use of numerical simulations
\citep{MCRS_howe2009, MCRS_tripathy2013}.

There are, however, standing problems affecting not only ring analysis but also
other local techniques.
Although 
the limiting low spatial resolution given by the selected region size
in comparison with time-distance analysis (and similar methods),
ring-diagram analysis is a well stablished and, in some respects, easy to interpret 
which makes it an important comparison tool
to solve the remaining problems and
fully understand the physical processes in the sun.

Much progress has being achieved and more is likely to come.
In the future, we hope to see improvements in several areas, such as:
the variation of the meridional flow with depth and with solar cycle;
the location of `counter cells' in the meridional circulation;
the formation, emergence , and decay of sunspots (i.e., their entire life-cycle); and 
how they drive and control the solar cycle.
We should always keep in mind that the Sun is an important reference for stellar physics,
and helioseismology a guide to asteroseismology.

\acknowledgements The author would like to thank CNPq/Brazil for partial travel support.


\bibliography{rabellosoares}

\end{document}